\documentclass[12pt]{article}
\usepackage{epsfig}

\newcommand{\be}{\begin{equation}}
\newcommand{\bea}{\begin{eqnarray}}
\newcommand{\ba}{\begin{array}}
\newcommand{\bean}{\begin{eqnarray*}}
\newcommand{\ee}{\end{equation}}
\newcommand{\eea}{\end{eqnarray}}
\newcommand{\ea}{\end{array}}
\newcommand{\eean}{\end{eqnarray*}}

\def\sigeff{\sigma_{\rm eff}}
\def\sigDS{\sigma_{\rm DS}}
\def\cO{{\cal O}}
\def\cL{{\cal L}}
\def\cR{{\cal R}}
\def\rar{{\rightarrow}}
\def\ds{double scattering }
\def\ss{single scattering }

\begin{document}

\begin{titlepage}

\begin{flushright}
DTP/99/110\\
hep-ph/9912232\\
December 1999
\end{flushright}
 
 \vspace*{3cm}
 
\begin{center}

{\Large\bf Like-Sign W Boson Production at the LHC \\ [3mm] as a Probe 
of Double Parton Scattering}\\[1.0cm]

{\large Anna~Kulesza$^1$  and  W.~James~Stirling$^{1,2}$}\\[0.5cm]
{\em $^1\ $Department of Physics, University of Durham, Durham, DH1 3LE, U.K.}
\\
{\em $^2\ $Department of Mathematical Sciences, University of Durham, Durham, 
DH1 3LE,  U.K.}
\end{center}
 
\vskip1.0cm    
\centerline{\bf   ABSTRACT}  
\vskip0.7cm    

\noindent 
Double parton scattering, i.e. two parton hard scattering processes
in the same hadron-hadron collision, may constitute an important background
for Higgs and other new particle searches at the LHC. We point out that
like-sign $W$ pair production provides a relatively clean way of searching
for, and calibrating, double parton scattering at the LHC.

\end{titlepage}

\noindent Recently~\cite{DFT} the importance of 
double parton scattering at the Large
Hadron Collider (LHC) has been readdressed. In particular it has been pointed
out~\cite{DFT} that double parton scattering may constitute 
a significant background to
Higgs boson production and decay via the $b\bar{b}$ decay channel, which, for
a Higgs mass below the $W^+W^-$ threshold, is one of the most promising
discovery channels. Of course, \ds contributes to the background in many
other processes, and similar analyses have been performed in the past
for hadron collisions at lower energy~\cite{DS}.
However, the LHC and its discovery potential necessitates a very accurate  
estimation of backgrounds where \ds may provide a significant contribution.
Therefore it
is essential to obtain a better quantitative understanding of double parton
scattering and a more precise estimation of the effect.

The double (multiple) scattering occurs when two (many) different pairs of partons scatter independently in
the same hadronic collision. From the theoretical point of view,
the presence of \ds is
required to preserve unitarity in the high energy limit, i.e. when the
distribution of partons with small momentum fractions within a hadron is high.  
In principle, \ds probes  correlations between partons in the hadron in the 
transverse plane,  and thus
provides important additional information on hadron structure~\cite{CT}. 
If  a scattering event is characterized by high centre-of-mass energy
and  relatively modest partonic subprocess energy,  which happens 
for example in the
production of heavy gauge bosons or a Higgs boson at the LHC, then 
parton-parton correlations can be assumed to be negligible. 
Such an assumption leads 
to a simple factorised expression for the \ds cross section 
(in the case of two distinguishable interactions, $a$ and $b$)~\cite{CT}   
\be
\sigDS ={{\sigma}_a {\sigma}_b \over \sigeff}\,.
\label{sigmaDS_d}
\ee
Here ${\sigma_a}$ represents  the single scattering cross section
\be 
{\sigma_a}=\sum_{i,j}\int d x_A d x_B f_i(x_A) f_j(x_B) 
\hat{\sigma}_{ij \to a}\,,
\label{sigmaSS}
\ee
with $f_i(x_A)$ being the standard parton distribution of parton $i$
and $\hat{\sigma}_{ij\to a}$ 
representing the partonic cross section.
If the two interactions are indistinguishable, double counting is avoided
by replacing Eq.~(\ref{sigmaDS_d}) with
\be
\sigDS ={{\sigma}_a {\sigma}_b \over 2 \sigeff} \,.
\label{sigmaDS_id}
\ee
The parameter $\sigeff$, the effective cross section, contains all the
information about the non-perturbative structure of the proton in this 
simplified approach and corresponds to the overlap of the matter
distributions in the colliding hadrons. The factorisation hypothesis appears 
to be in
agreement with the experimental data from CDF~\cite{CDF} at
the Tevatron $p \bar p $ collider. It is also
believed that $\sigeff$ is largely independent of the centre-of-mass
 energy of the
collision and on the nature of the  partonic interactions (for a detailed
discussion the reader is referred to~\cite{CT}). Therefore throughout this
study we will use the value $\sigeff=14.5~{\rm mb}$, as measured by CDF.~%
\footnote{Strictly speaking, this value refers to an exclusive measurement
 and therefore should be understood as an {\it upper} bound on $\sigeff$.}

Given the potential importance of double scattering as a background to
new-physics searches at the LHC, it is important to be able
to calibrate the effect, i.e. to measure $\sigeff$ using a known,
well-understood Standard Model process. In the case of single scattering
processes, the benchmark process is $W$ boson production, see for example
Ref.~\cite{MRST99}.~\footnote{It has 
even been suggested that this process could be used to measure the luminosity 
at the LHC.} This suggests that $W$ {\it pair} production could be used
to calibrate double parton scattering. In the Standard Model, like-sign
$W$ pair production is much smaller than opposite-sign production, which
suggests that the former channel is the best place to look for additional 
double scattering contributions.

The purpose of this note is to quantify the expected cross sections
for like- and opposite-sign $W$ pair production at the LHC, from both the 
single and double scattering mechanisms, and to explore differences
in the distributions of the final state particles.

The predicted rate of single $W$ production at the LHC is naturally very
high, resulting in a significant  \ds cross section. Since the 
$W^+$ and $W^-$ \ss cross sections are comparable in magnitude, the same will
be true for the double scattering $\sigDS(W^+W^+)$, $\sigDS(W^-W^-)$ and 
$\sigDS(W^+W^-)$ cross sections. However, for single scattering
we would expect  \mbox{$\sigma(W^-W^-)< \sigma(W^+W^+) \ll \sigma(W^+W^-)$}.
The reason is that while the latter is $\cO(\alpha_W^2)$ at leading order,
same-sign  inclusive $W$ pair production 
is a mixed
strong-electroweak process with leading contributions of 
$\cO(\alpha_S^2\alpha_W^2)$ and $\cO(\alpha_W^4)$. Hence we might
expect that  like-sign $W$  pair  production,  with its relatively larger 
\ds component,  could give a  clean
measurement of $\sigeff$. 

The possibility of \ds `background' contributions to  like-sign
$W$ pair production  was noticed some time
ago~\cite{BCHP}, when this process was considered as 
one of the most promising channels for searching for
 strong scattering in the
electroweak symmetry breaking sector~\cite{SSB}. In these  studies
 the \ds contribution was treated as an unwanted
background and suppressed by applying appropriate cuts. 
 
We begin our analysis by calculating the total single-scattering 
cross sections for single $W$
and (opposite-sign and like-sign) $W$ pair production in $pp$ and $p \bar p$
collisions at scattering energy $\sqrt{s}$. 
For consistency, we consider only {\it leading-order}
 cross sections for all processes studied, i.e. we use leading-order
subprocess cross sections with leading-order parton 
distributions.~\footnote{We note that  the 
full $\cO(\alpha_S^2)$ corrections to single $W$~\cite{WNNLO} and
$\cO(\alpha_S)$ corrections to $W$ pair production~\cite{WWNLO} have been calculated.}

\begin{figure}[htp]
\begin{center}
\mbox{\epsfig{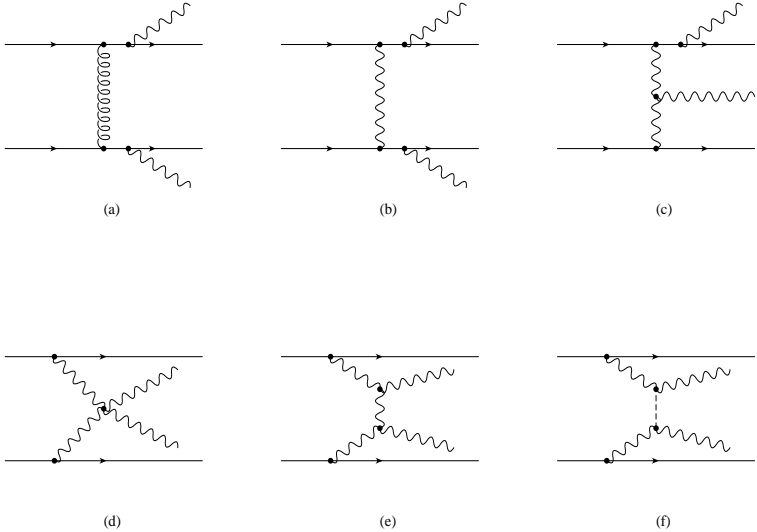}}
\caption{Examples of Feynman diagrams for the $uu \to W^+W^+dd$
scattering process, ${\cal O}(\alpha_S^2 \alpha_W^2)$ (a) and ${\cal O}(\alpha_W^4)$ (b-f).}
\end{center}
\end{figure}
%
As already noted, in the context of  leading-order single parton 
scattering, opposite-sign $W$ pair
production in hadron-hadron collisions arises from the  $\cO(\alpha_W^2)$ subprocess
\be
q  + \bar {q} \to W^+ + W^-
\ee
In contrast, like-sign $W$ pair production is an
$\cO(\alpha_S^2\alpha_W^2)$ or $\cO(\alpha_W^4)$ process at leading order:
\be
q + q\to W^+ + W^+ + q' +  q' 
\ee
with $q=u,c,\ldots$, $q'=d,s,\ldots$, together with the corresponding
crossed processes. Charge conjugation gives a similar set of subprocesses for
$W^-W^-$ production.
The Feynman diagrams split into two groups: the first
set corresponds to the  $\cO(\alpha_S^2\alpha_W^2)$  gluon exchange process $qq \to qq$
where a single $W$ is  emitted from each of the quark lines, see
Fig.~1(a). The second, $\cO(\alpha_W^4)$, set contains analogous electroweak
diagrams,  i.e.  $t-$channel $\gamma$ or $Z$ exchange,  as well as $WW$
scattering diagrams, including also a $t-$channel Higgs exchange contribution,
see Fig.~1(b-f).
Note that the corresponding cross sections are infra-red and 
collinear safe: the total rate can be calculated without imposing any
cuts on the final-state quark jets. We would therefore expect naive
coupling constant power counting to give the correct order of
magnitude difference between  the like-sign and opposite-sign cross 
sections, i.e. $\sigma(W^+W^+)\; \sim\; \alpha_{S,W}^2\; \sigma(W^+W^-)$.
Given the excess of $u$ quarks over
$d$ quarks in the proton, we would 
also expect $\sigma(W^+W^+)\; >\; \sigma(W^-W^-)$.

\begin{figure}[htp]
\begin{center}
\mbox{\epsfig{figure=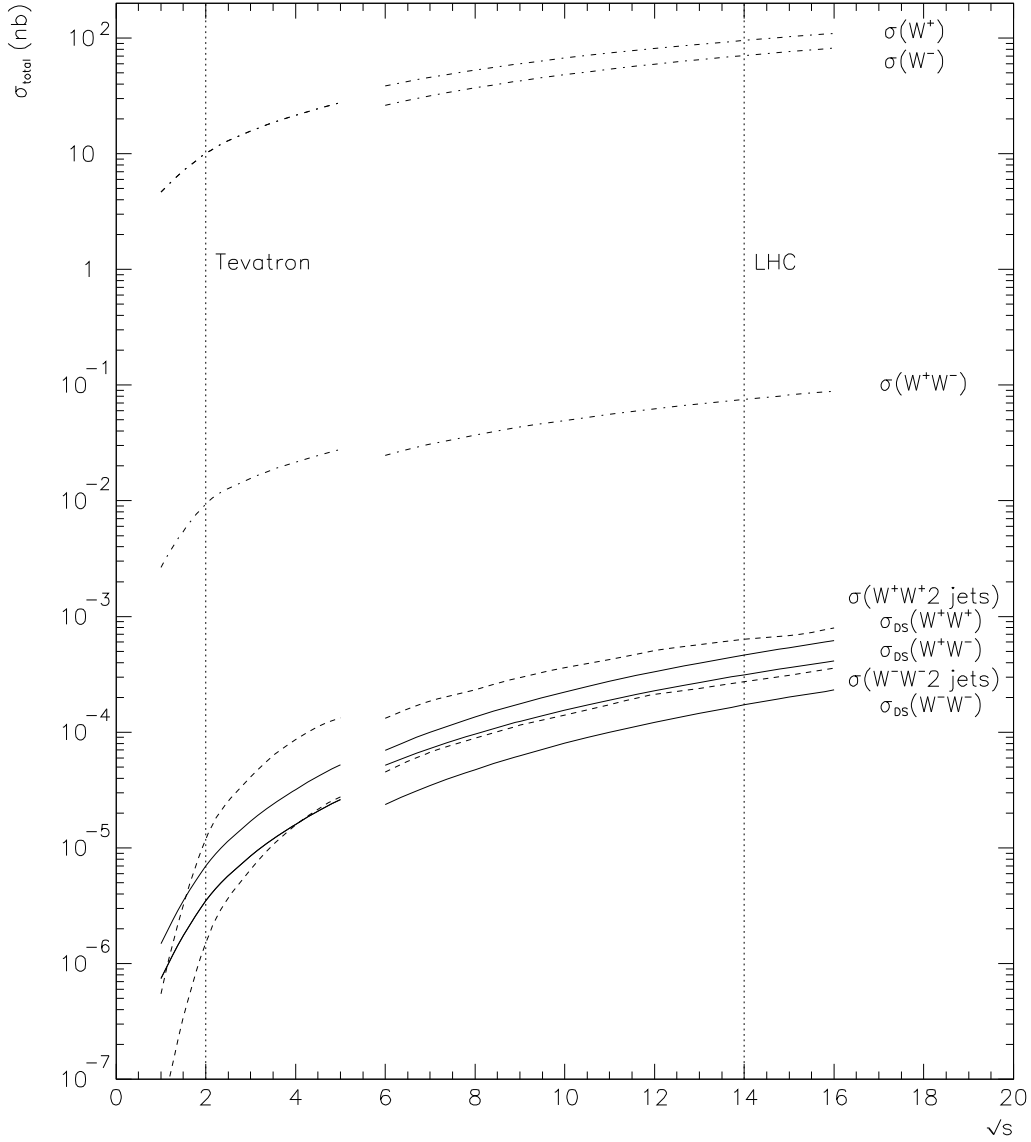,width=15.0cm}}
\caption{Total cross sections for single $W$ and $W$ pair production in
$pp$ and $p\bar{p}$ collisions. The dashed and dot-dashed lines correspond to
single parton scattering, and the solid lines to double parton scattering assuming
$\sigeff = 14.5$ mb.}
\end{center}
\end{figure}

Figure~2 shows the total single $W$ and $W$ pair 
cross sections in  proton--antiproton and proton--proton collisions as a 
function of the collider energy. No branching ratios are included,
and there are no cuts  on any of the final state particles. The matrix
elements are obtained using MADGRAPH~\cite{MADGRAPH} and HELAS~\cite{HELAS}. 
We use the 
MRST leading-order parton distributions from Ref.~\cite{MRST}, and the most
recent values for the electroweak parameters.~\footnote{Note that the same-sign
cross sections are weakly dependent on the Higgs mass: varying the mass from
 $M_H=125$~GeV to $M_H=150$~GeV leads to only a 
 2\% change in the total rate at the LHC. We use $M_H=125$~GeV as the default value.} 
 Note that 
for $p \bar{p}$ collisions, $\sigma(W^+) \equiv \sigma(W^-)$
and $\sigma(W^+W^+) \equiv \sigma(W^-W^-)$. The like-sign and opposite-sign
cross sections differ by about two orders of magnitude, as expected.
Despite the fact that $\alpha_S > \alpha_W$, 
 the electroweak contribution to the \ss
like-sign $WW$ production cross section 
is similar in size to the strong contribution.
This is due to the relatively large number of diagrams
 (e.g. 68 for $u u \rar W^+ W^+ d
d$), as compared to the gluon exchange contribution (16 for the same process).
A total annual luminosity of $\cL = 10^5$~pb$^{-1}$ at the LHC
would yield approximately  $65$ thousand $W^+W^+$ events and
$29$ thousand $ W^-W^-$ events, before
high-order corrections, branching ratios
 and acceptance cuts are included.

The production characteristics of the $W$s in like- and opposite-sign
production  are somewhat different. In particular, the presence 
of two jets in the final state for the former leads to a broader
transverse momentum distribution, as illustrated in Fig.~3. Also of interest
is the jet transverse momentum distribution in $W^\pm W^\pm$ production,
shown in Fig.~4. This indicates that a significant fraction of the jets
would pass a detection $p_T$ threshold, and could be used as an 
additional `tag' for like-sign production. Of course one also expects
large $p_T$ jets in opposite-sign $W$ production via higher-order processes,
e.g. $q \bar{q} \to W^+W^-g$ at $\cO(\alpha_S)$, but these have a 
steeply falling distribution reflecting the underlying
infra-red and collinear singularities at $p_T = 0$. 

\begin{figure}[htp]
\begin{center}
\mbox{\epsfig{figure=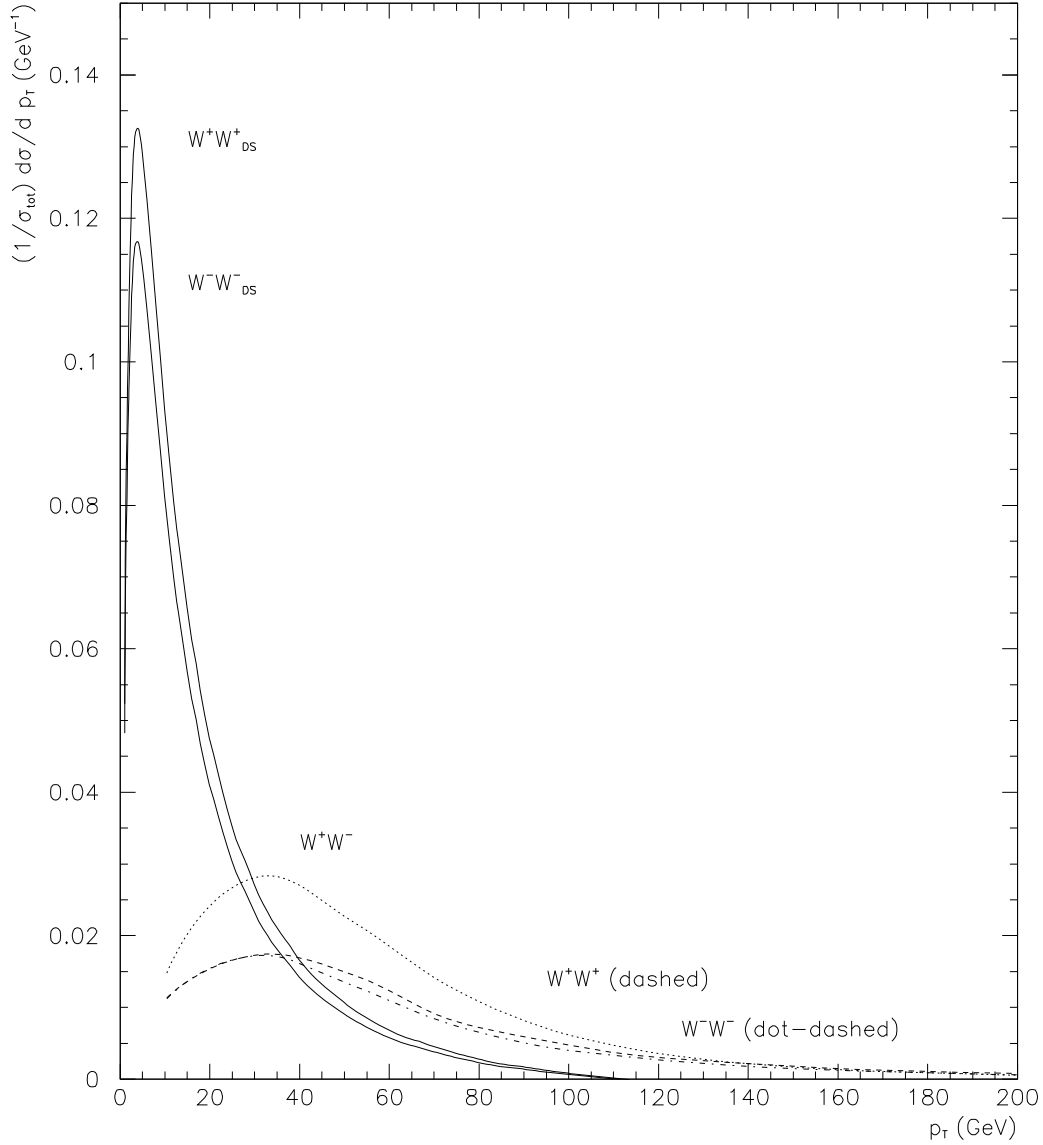,width=15.0cm}}
\caption{Transverse momentum distributions for $W^+W^+$, $W^-W^-$, $W^+W^-$
(dashed, dot-dashed and dotted lines, respectively) single
parton scattering and $W^+W^+$, $W^-W^-$ double parton scattering (solid
lines) at the LHC. The \ds predictions are obtained using the $p_T$-space resummation
method~\cite{KS} (with neither smearing nor matching included).}
\end{center}
\end{figure}

\begin{figure}[htp]
\begin{center}
\mbox{\epsfig{figure=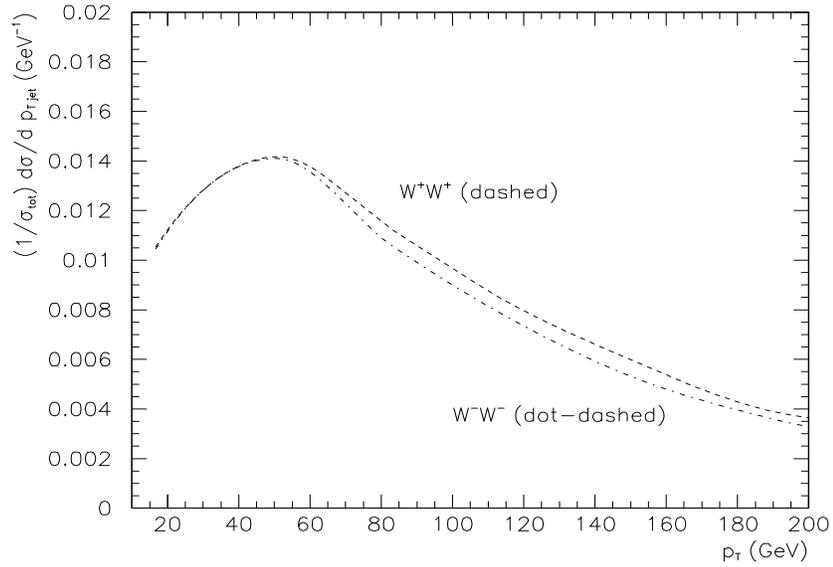,width=12.0cm,height=9cm}}
\caption{Jet transverse momentum distributions in like-sign
\ss $WW$ production.}
\end{center}
\end{figure}

\begin{figure}[htp]
\begin{center}
\mbox{\epsfig{figure=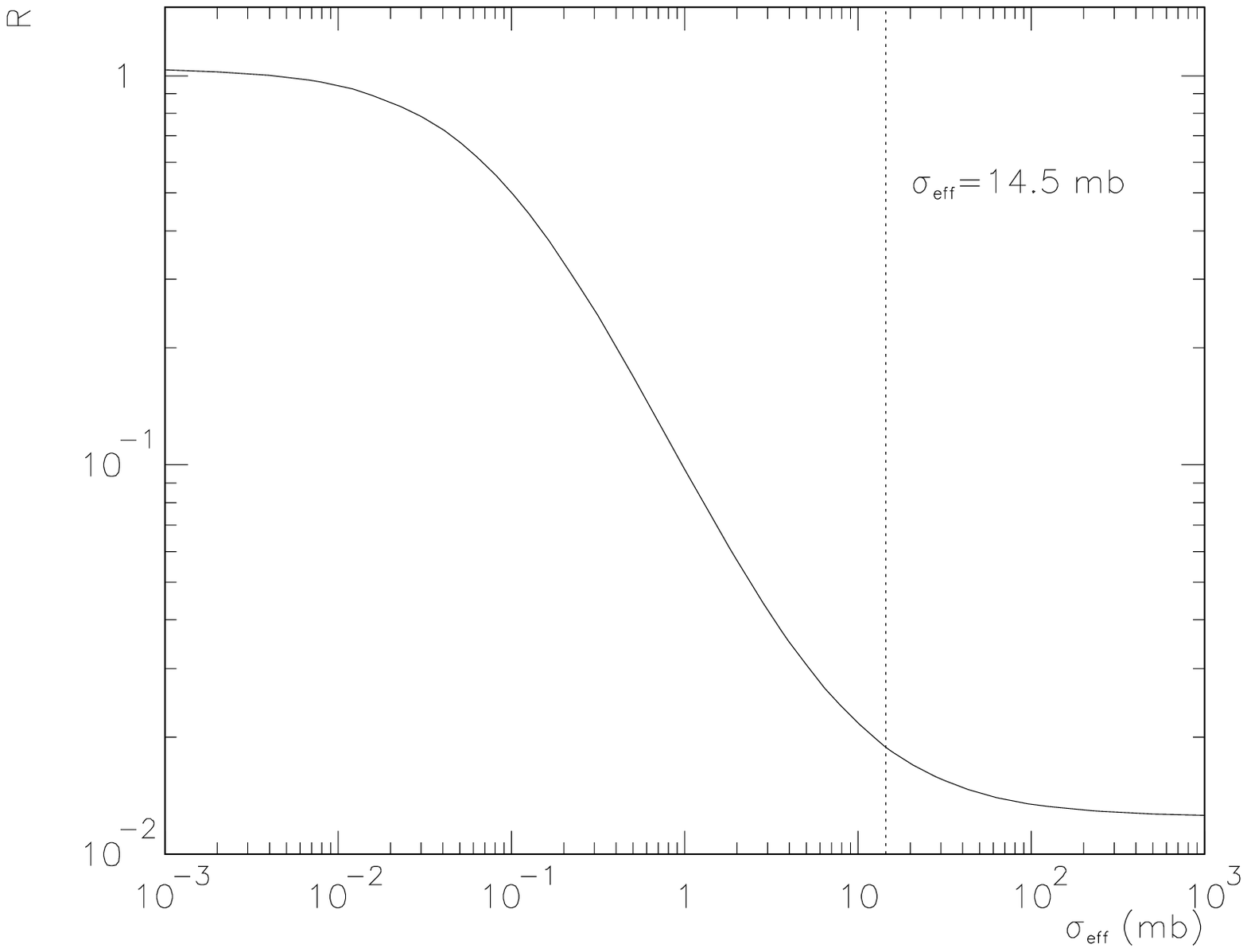,width=12.0cm,height=9cm}}
\caption{The dependence of the ratio $\cR$ of like-sign to opposite-sign
$W$ pair event rates on the effective cross section $\sigeff$ at the LHC.}
\end{center}
\end{figure}

We turn now to the double parton scattering cross sections.
As discussed above, we estimate these by simply multiplying the corresponding
\ss cross sections and normalising by $\sigeff$ for the
like-sign $W$ pair production and $2\sigeff$ for the opposite-sign case.
The factorisation assumption holds since the energy required to produce a
vector boson is much lower than the overall centre of mass energy. 
Figure~2 shows the resulting total $\sigDS(W^+W^-)$ and $\sigDS(W^\pm W^\pm)$
cross sections as a function of $\sqrt{s}$. The opposite-sign single
scattering and \ds cross sections differ by two orders of
magnitude. However for like-sign $W^+W^+$ ($W^-W^-$) 
production the \ds scattering contribution
is only a factor 2.1 (1.7) smaller than the \ss contribution.
Additionally, a \ds event
signature differs significantly from the \ss case.
In particular, the $W$ transverse momentum distribution from \ds has a
very pronounced, steep peak for small values of $p_T$ (see Fig.~3),
inherited from the \ss $p_T$ distribution~\footnote{We are assuming here that
the non-perturbative `intrinsic' transverse momentum distributions
of the two partons participating in the double parton scattering
are uncorrelated.}, in contrast to the broader single-scattering
distributions.  Obviously, similar
features will characterize the $p_T$ spectra of leptons originating from  $W$
decay, allowing for additional discrimination between double and single
scattering events. 

\begin{table}[tb]
\begin{center}
\begin{tabular}{|l|l|l|l|}\hline
 & $N(W^+W^-)$  & $N(W^+W^+)$ & $N(W^-W^-)$
    \\ \hline
single scattering  & \ \ 7,500,000  & \ \ \ \ 65,000 &  \ \ \ \ 29,000 \\
double scattering  & \ \ \ \ 46,000  & \ \ \ \ 31,000  & \ \ \ \  17,000 \\
\hline
\end{tabular}
\label{tab1}
\caption{The  expected number of $WW$ events expected for ${\cal L} 
= 10^5$~pb$^{-1}$ at the LHC
  from single
and double scattering, assuming $\protect\sigeff=14.5$~mb for the latter.}
\end{center}
\end{table}               
 
The absolute rate of like-sign $W^+W^+$ and $W^-W^-$ pair production therefore
provides a relatively clean  measure of $\sigeff$ at LHC energies. Table~1
summarizes the number of expected events in the various $WW$ channels (recall
these are leading-order estimates only, with no branching ratios), assuming  
$\sigeff=14.5$~mb. 
However, since the absolute event rates shown in Table~1  are
 sensitive to overall measurement and theoretical
uncertaintities, it may be more useful to consider cross section {\it ratios}.
Consider for example the ratio of the like- to opposite-sign event rates
\bea
\cR & =&  {N(W^+W^+) + N(W^- W^-) \over N(W^+W^-) } \nonumber \\
 &=& {\sigma(W^+W^+) + \sigma(W^- W^-) +(2 \sigeff)^{-1} \left[
 \sigma(W^+)^2 +  \sigma(W^-)^2 \right]  \over 
 \sigma(W^+W^-) + \sigeff^{-1} 
 \sigma(W^+) \sigma(W^-) }
\label{ratio}
\eea 
with both single and \ds contributions included. 
The ratio $\cR$ for the LHC  is shown as a function of $\sigeff$ 
in Fig.~5. The limit $\sigeff \to \infty$ corresponds to the 
(very small) \ss ratio, $\cR=0.0125$ while $\sigeff \to 0$
corresponds to the ratio ($\approx 1.05$) of the {\it single} $W$ production
cross sections in $pp$ collisions. The CDF measured value \cite{CDF}
of $\sigeff = 14.5$~mb gives $\cR = 0.019$.

In conclusion, we have shown that like-sign $W$ pair production 
provides a relatively clean environment for searching for and
calibrating double parton scattering at the LHC. A measurement of $\sigeff$ from
this process would allow the double scattering backgrounds to 
new physics searches to be calibrated with precision.
In this brief study we have concentrated on overall total event rates. 
An interesting next step would be to perform more detailed
Monte Carlo studies of the various production processes, taking into
account the $W$ decays, experimental acceptance cuts, etc. 
In fact it would not be difficult to devise additional cuts to enhance
the double scattering contribution. We see from Fig.~3, for example,
that a cut of $p_T(W) < {\cal O}(20~{\rm GeV})$ would remove most of the 
single scattering events while leaving the double scattering
contribution largely intact.

\vskip 1truecm

\noindent{\bf Acknowledgements}
\medskip

\noindent 
This work was supported in part by the EU Fourth Framework Programme
`Training and Mobility of Researchers', Network `Quantum Chromodynamics and
the Deep Structure of Elementary Particles', contract FMRX-CT98-0194 (DG 12 -
MIHT). A.K. gratefully acknowledges financial support received from the 
ORS Award Scheme and the University of Durham.  
\goodbreak


\begin{thebibliography}{99}

\bibitem{DFT} 
A.~Del Fabro and D.~Treleani, preprint hep-ph/9911358.
\bibitem{DS}
P.V.~Landshoff and J.C.~Polkinghorne, {\it Phys. Rev.} {\bf D18} (1978) 3344; \\
C.~Goebel, F.~Halzen and D.M.~Scott, {\it Phys. Rev.} {\bf D22} (1980) 2789;\\
N.~Paver and D.~Treleani, {\it Nuovo Cimento} {\bf A70} (1982) 215; \\
F.~Halzen, P.~Hoyer and W.J.~Stirling, {\it Phys. Lett.} {\bf B188} (1987) 375; \\
M.~Mangano, {\it Z. Phys.} {\bf C42} (1989) 331;  \\
R.M.~Godbole, S.~Gupta and J.~Lindfors, {\it Z. Phys.} {\bf C47} (1990) 69;  \\
M.~Drees and T.~Han, {\it Phys. Rev. Lett.} {\bf 77} (1996) 4142.
\bibitem{CDF}
F.~Abe et al. (CDF Collaboration), {\it Phys. Rev. Lett.} {\bf 79} (1997) 584;
{\it Phys. Rev.} {\bf D56} (1997) 3811.
\bibitem{BCHP}
V.~Barger, Kingman~Cheung, T.~Han and R.J.N.~Phillips, {\it Phys. Rev.} {\bf
  D42} (1990) 3052. 
\bibitem{SSB}
D.A.~Dicus and R.~Vega, {\it Phys. Lett.} {\bf B217} (1989) 194; 
 {\it Nucl. Phys.} {\bf B329}  (1990) 533; \\
M.S.~Chanowitz and M.~Golden, {\it Phys. Rev. Lett.} {\bf 61} (1988) 1053; \\ 
M.S.~Berger and M.S.~Chanowitz, {\it Phys. Lett.} {\bf B263} (1991) 509;  \\ 
D.A.~Dicus, J.F.~Gunion, L.H.~Orr and R.~Vega, {\it Nucl. Phys.} {\bf
  B377} (1991) 31.
\bibitem{CT}
G.~Calucci and D.~Treleani, {\it Phys. Rev.}  {\bf D57} (1998) 503; 
{\it ibid.} {\bf D60} (1999) 054023. 
\bibitem{MRST99}
A.D.~Martin, R.G.~Roberts, W.J.~Stirling and R.S.~Thorne, 
preprint hep-ph/9907231.
\bibitem{WNNLO}
R.~Hamberg, T.~Matsuura and W.L.~van~Neerven, {\it  Nucl. Phys.}  
{\bf B345} (1990) 331; 
{\it ibid.} {\bf B359} (1991) 343; \\   
W.L.~van~Neerven and E.B.~Zijlstra, {\it Nucl. Phys.}  {\bf B382} (1992) 11.
\bibitem{WWNLO}
J.~Ohnemus, {\it Phys. Rev.} {\bf D44} (1991) 1403;\\
S.~Frixione, {\it Nucl. Phys.} {\bf B410} (1993) 280; \\
J.~Ohnemus, {\it Phys. Rev.} {\bf D50} (1994) 1931;\\  
L.~Dixon, Z.~Kunszt and A.~Signer, 
{\it Nucl. Phys.} {\bf B531} (1998) 23; 
{\it Phys. Rev.} {\bf D60} (1999) 114037;\\
J.M.~Campbell and  R.K.~Ellis, {\it Phys. Rev.} {\bf D60} (1999) 113006. 
\bibitem{MADGRAPH}  
T.~Stelzer and W.F.~Long, {\it Comp. Phys. Comm.} {\bf 81} (1994) 357. 
\bibitem{HELAS} 
E.~Murayama, I.~Watanabe and K.~Hagiwara, {\it KEK report} 91-11, January
1992.
\bibitem{VEGAS}
G.P.~Lepage, {\it Jou. Comp. Phys.} {\bf 27} (1978) 192.
\bibitem{MRST}
A.D.~Martin, R.G.~Roberts, W.J.~Stirling and R.S.~Thorne, 
Eur. Phys. J. {\bf C4} (1998)  463.
\bibitem{KS}
A.~Kulesza and W.J.~Stirling, in preparation.

\end{thebibliography}
\end{document}